\documentclass[%
twocolumn,
 amsmath,amssymb,
 aps,
floatfix,
]{revtex4-2}

\usepackage{graphicx}
\usepackage{dcolumn}
\usepackage{bm}
\usepackage{physics}
\usepackage{subcaption}
\usepackage[english]{babel}
\usepackage{tabularx}
\usepackage{hyperref}
\hypersetup{
    colorlinks=true,
    linkcolor=blue,
    citecolor=red,      
    urlcolor=blue,
    pdftitle={draft},
    }
    
\begin{document}

\def\be{\begin{equation}}
\def\ee{\end{equation}}

\setlength{\parskip}{0.2\baselineskip}

\title{Optimal Quantum Feshbach Engines}

\author{Aaron Wandhammer}
\author{Vincent Hardel}
\author{Paul-Antoine Hervieux}
\author{Giovanni Manfredi}
\email{giovanni.manfredi@ipcms.unistra.fr}
\affiliation{University of Strasbourg, CNRS, Institut de Physique et Chimie des Mat\'eriaux de Strasbourg, UMR 7504, F-67034 Strasbourg, France
}

\date{\today}

\begin{abstract}
We develop an optimization framework for high-efficiency quantum cycles implemented with a trapped Bose-Einstein condensate, whose control parameters are the trap stiffness and the interaction strength tuned via a Feshbach resonance.
Optimal driving protocols for each stroke of the cycle are obtained from a variational description of the condensate dynamics combined with Nelson's stochastic quantization, which maps the quantum evolution onto an effective Ornstein-Uhlenbeck process. 
The optimal protocol is obtained by minimizing a user-defined cost functional that selects the best trade-off between protocol duration and arbitrary physical constraints (such as the expended work or the proximity to an adiabatic evolution), and exhibits remarkable stability over repeated cycles. 
The method also provides a systematic route to optimal control for  generic nonlinear Schr\"odinger equations, paving the way to optimal control strategies in fields as diverse as nonlinear optics, quantum fluids, and quantum plasmas.
\end{abstract}

\maketitle


{\it Introduction.}--- The precise control of quantum matter is a central challenge in modern  quantum physics, as it underpins the development of emerging quantum technologies \cite{cirac_goals_2012,deffner_quantum_2019}. 
The most reliable way to transfer a quantum system between states in a stable manner is through adiabatic transitions. Such transitions, however, require theoretically infinite time to maintain perfect stability throughout the evolution. Consequently, adiabatic protocols are often impractically slow for experimental applications. 
To overcome this limitation, the field of ``Shortcuts to Adiabaticity" (STA) has emerged \cite{torrontegui_shortcuts_2013,GueryOdelin2019}. 
STA encompasses a broad set of techniques --- ranging from inverse engineering \cite{li_shortcut_2016,torrontegui_fast_2011} and counter-adiabatic driving \cite{del_campo_shortcuts_2013,berry_transitionless_2009} to fast-forward approaches \cite{Masuda2010} --- which enable finite-time transitions while reproducing the outcome of an adiabatic evolution. These methods provide experimentally robust protocols that steer the quantum system towards a stable final state by
manipulating some control parameter.  Despite two decades of progress, STA remains a topical area of research \cite{plata_taming_2021,raynal_shortcuts_2023}.

Recently, a novel STA technique was introduced \cite{hardel_shortcuts_2024}, based on a quantum-classical analogy derived from Nelson’s formulation of quantum mechanics \cite{nelson_derivation_1966}, which 
recasts the Schr\"odinger equation as a classical stochastic process.
Here, we use this approach to control the dynamics of a Bose--Einstein condensate (BEC) confined in a cigar-shaped harmonic trap \cite{castin_bose-einstein_2001,dalfovo_theory_1999}. 
For ultracold atomic gases, the interaction strength is characterized by the scattering length $a$, which can be tuned with an external magnetic field that acts on the Feshbach resonance \cite{kagan_evolution_1997,roati_k_2007,derrico_feshbach_2007,chin_feshbach_2010}. This tunability provides an additional control parameter, alongside the trap stiffness \cite{del_campo_frictionless_2011}, for driving quantum state transitions. 
STA methods tailored to BECs with tunable interactions thus hold significant promise, notably in the development of quantum engines based on ultracold atomic gases \cite{Schaff2011,Brandner2015,Beau2016,FengWu2006,li_efficient_2018,keller_feshbach_2020,Nautiyal2024,Estrada2024,Mishra2025}.

Using Nelson's stochastic quantization, we first derive faster-than-adiabatic quantum protocols inspired by classical protocols originally developed for underdamped stochastic systems. We then develop optimal quantum protocols and optimal quantum cycles (engines) by exploiting the tunability of both the trap stiffness and the scattering length of the BEC.

Compared to other STA techniques \cite{Schaff2011,Mishra2025}, our method enables the construction of protocols that are optimal with respect to a cost functional freely specified by the user. This flexibility allows one to identify the best trade-off between, for instance, the protocol duration, the expended work, or the proximity to an adiabatic evolution. Under realistic conditions, this approach leads to a significant increase in the efficiency and generated power of a quantum engine based on a BEC.

{\it Model.}--- We consider a BEC confined in an elongated harmonic trap, where the longitudinal trapping frequency (along the $x$-axis) is much smaller than the transverse one, $\omega_x \equiv \omega \ll \omega_\perp$, making the system quasi-one-dimensional (1D) \cite{fort_effect_2005,Manfredi2008}. In the mean-field approximation, the evolution of the condensate is governed by the Gross-Pitaevskii equation (GPE)
\begin{equation}\label{eq:GP}
    i\hbar \frac{\partial \psi}{\partial t}= \left( -\frac{\hbar^{2}}{2m}\laplacian{} + \frac{1}{2}m\omega^{2}x^{2} + g_{\rm 1D}N \abs{\psi}^{2} \right) \psi,
\end{equation}
where $\psi(x,t)$ is the condensate wave function, $g_{\rm 1D}=2a \hbar\omega_{\perp}$ is the effective 1D coupling constant, $a$ is the 3D scattering length, and $N$ is the number of atoms in the BEC. For simplicity, hereafter we write $g\equiv g_{\rm 1D}$ and define the trap stiffness as $\kappa \equiv m\omega^2$.
Here, the purpose is to steer the evolution of the BEC by acting on the stiffness of the trap $\kappa(t)$ and/or the interaction strength $g(t)$.

To obtain an approximate solution of the GPE, we write the wave function as $\psi(x,t)=\sqrt{P(x,t)}\, e^{iS(x,t)}$, and assume that the density $P$ is Gaussian and the phase $S$ is quadratic in $x$, i.e.: $P(x,t)=\frac{1}{\sqrt{2\pi}\sigma(t)}\, e^{-x^2/2\sigma^2(t)}$ and $S(x,t)=\alpha(t) x^{2}$.
From the continuity equation for the density $P(x,t)$, it can be deduced that $\alpha(t)=\frac{m}{2\hbar}\frac{\dot \sigma}{\sigma}$, where the dot denotes time differentiation. The mean-squared displacement $\sigma(t)$ obeys an Ermakov equation that can be obtained using variational arguments \cite{Haas2009,Hurst2016} (see Appendix \ref{app:variational} for details). 
When inserted into the Gaussian ansatz for $\psi$, this procedure yields the Gaussian wave function that best approximates the exact solution of the GPE.
This approximation is exact for $g=0$ and remains reasonably accurate for moderate interactions, as was shown for a Coulomb gas \cite{Haas2009}. 


{\it Quantum-classical analog.}--- We now introduce the quantum-classical analogy based on Nelson's stochastic quantization \cite{nelson_derivation_1966,Bacciagaluppi1999}. The latter stipulates that the Schrödinger equation is equivalent to a classical stochastic process governed by a first-order Langevin equation
$ \mathrm{d}x(t) = b(x(t),t)\mathrm{d}t + \mathrm{d}{W}$,
where $\mathrm{d} x$ is the position increment, $\mathrm{d}t$ is the time step, and $\mathrm{d}W$ is a Wiener process with zero mean and correlations $\langle \mathrm{d}W(t) \mathrm{d}{W}(t') \rangle = 2D \mathrm{d}t\,\delta(t-t')$, with $D = \hbar/2m$ being the quantum diffusion coefficient.
The drift velocity $b(x,t)$ can be written in terms of the amplitude and phase of the wave function,  $b(x,t)=\frac{\hbar}{m}\partial_x S(x,t) + D\partial_x{\ln{P(x,t)}}$.

When the wave function is Gaussian, as we have assumed earlier, the drift velocity is linear in $x$, and the Nelson stochastic equation can be written as
\begin{equation}\label{eq:OUquantum}
    \mathrm{d}x=\frac{\hbar}{m}\left(2\alpha(t)-\frac{1}{2\sigma^{2}(t)}\right)x~\mathrm{d}t +\mathrm{d}W.
\end{equation}
The above equation is identical to an overdamped Ornstein-Uhlenbeck (OU) process \cite{uhlenbeck_theory_1930}:
$\mathrm{d}x = -\frac{\bar{\kappa}(t)}{\gamma}x \mathrm{d}t +\mathrm{d}W$,
where $\bar{\kappa}(t)$ is a the stiffness of a classical harmonic trap and $\gamma$ is a drag coefficient that we include for dimensional consistency. 
The classical OU equation and the quantum Nelson equation become identical when the coefficients in front of the first term on the right-hand side are the same. Equating these two coefficients, one finds, after some algebra (see Appendix \ref{app:bridge})
\begin{equation}\label{eq:bridge}
    \kappa(t)-\frac{Ng(t)}{4\sqrt{\pi}\sigma^{3}(t)}=\frac{\hbar^{2}}{2m\sigma^{4}(t)}+\frac{m}{\gamma}\dot{\bar{\kappa}}(t) - \frac{m}{\gamma^{2}}{\bar{\kappa}}^{2}(t),
\end{equation}
which is the central element of the quantum-classical analogy \cite{hardel_shortcuts_2024}.

The variance $s(t)$ of the OU process obeys the equation  
\begin{equation}\label{eq:variance}
\frac{d s}{dt} = \frac{2}{\gamma}\,[D\gamma-\bar{\kappa}(s)\,s],
 \end{equation}
with equilibrium condition $\bar{\kappa}_{eq}= D\gamma/s_{eq}$.
In virtue of Eq. \eqref{eq:bridge}, the variance $s$ is identical to the squared width $\sigma^2$ of the Gaussian wave function: $s(t) = \sigma^2(t)$. 
In other words, the approximate Gaussian solution of the GPE with time-dependent stiffness $\kappa(t)$ and interaction strength $g(t)$ is identical to the solution of a classical OU process with oscillator stiffness ${\bar \kappa}(t)$, provided  $\kappa$, $g$ and $\bar \kappa$ satisfy Eq. \eqref{eq:bridge}.
Hence, a BEC for which either $g(t)$ or $\kappa(t)$ (or both) are time-dependent can be emulated by a classical stochastic process with variable stiffness ${\bar \kappa}(t)$.

In practice, in order to obtain the quantum protocol, one proceeds in three steps: (i) fix the classical stiffness $\bar\kappa(t)$, (ii) solve Eq. \eqref{eq:variance} to obtain the variance $s(t)$, and (iii) use Eq. \eqref{eq:bridge} to obtain either $\kappa(t)$ at fixed $g$ or $g(t)$ at fixed $\kappa$. Finally, the GPE is solved numerically~\footnote{We use a standard finite-difference scheme (Crank-Nicolson with split operator between the kinetic and potential parts of the Hamiltonian). The initial condition is computed self-consistently and not assumed to be Gaussian.} with the obtained quantum protocol $g(t)$ or $\kappa(t)$ and $\langle x^2 \rangle = \int x^2 |\psi|^2 \mathrm{d}x$ is compared to the expected variance $s(t)$.

\begin{figure}
\includegraphics*[width=1\linewidth]{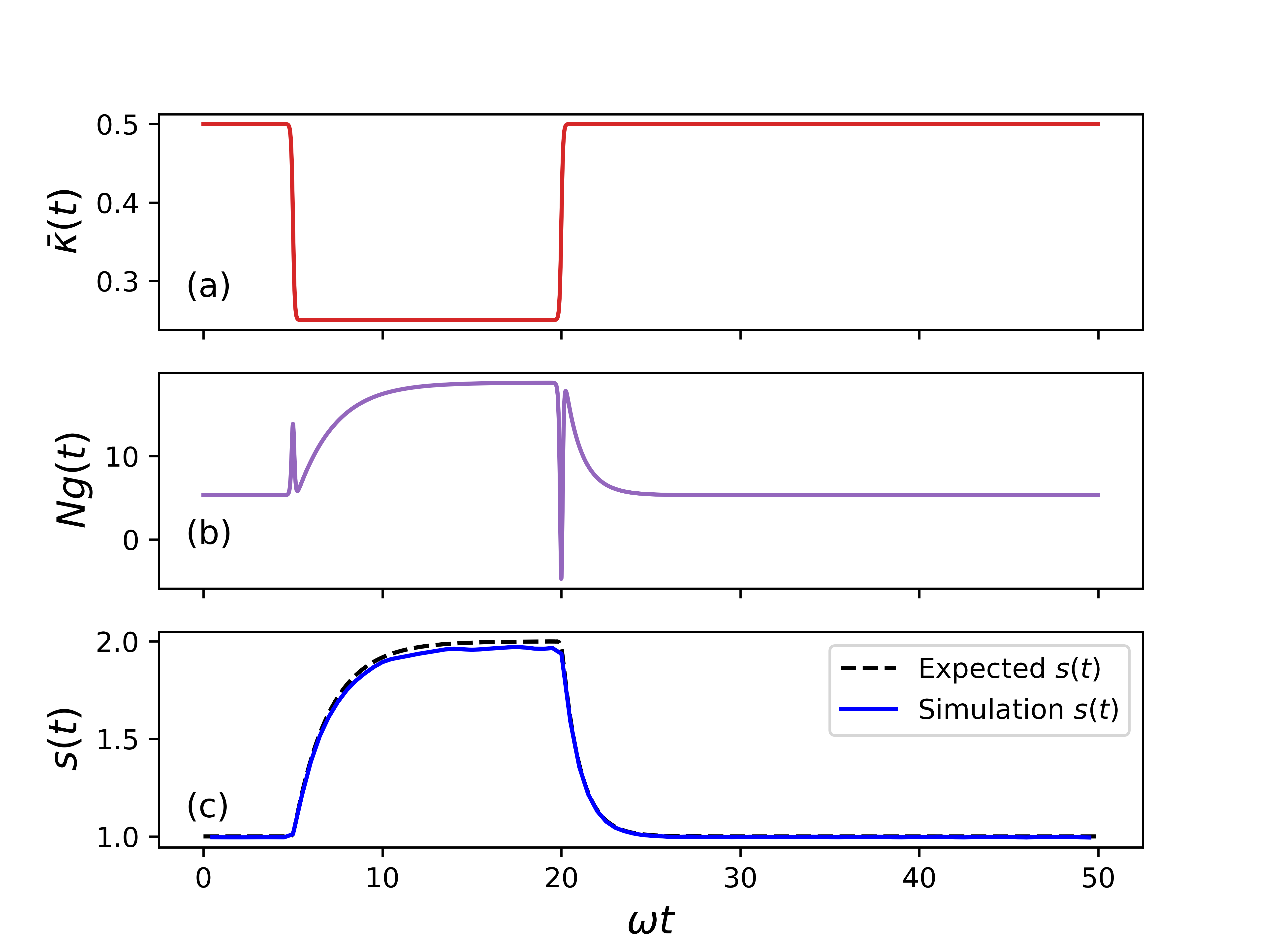}
    \caption{Steplike protocol on the classical stiffness ${\bar\kappa}(t)$ (a) and corresponding quantum $g(t)$ protocol computed from Eq. \eqref{eq:bridge} (b). The bottom panel (c) shows the variance $s(t)$ obtained from Eq. \eqref{eq:variance} (dashed black line) and that computed from a numerical simulation of the GPE (solid black line) using the time-dependent interaction strength $g(t)$ shown in panel (b). Here, $s$ is expressed in units of $L^2_{ho}=\hbar/(m\omega)$, $\bar \kappa$ in units of $m\omega^2$, and $g$ in units of $\hbar \omega L_{ho}$.}
\label{fig:STEP}
\end{figure}

{\it Shortcut to adiabaticity.}--- As a first example, we assume a smoothed step function for the classical stiffness $\bar\kappa(t)$, as represented in Fig. \ref{fig:STEP}(a). Using Eq. \eqref{eq:bridge} with $\kappa$ constant, we obtain the functional form of the interaction strength $g(t)$, shown in Fig. \ref{fig:STEP}(b). Such a form is highly nontrivial, but nicely ensures the transition from $s_i=1$ to $s_f=2$ and backwards, in a time of the order of $15\, \omega^{-1}$, i.e., about 2.5 oscillation periods of the trap potential---a significant improvement compared to an adiabatic (infinitely slow) protocol.
Directly injecting this protocol $g(t)$ into the GPE yields a result very similar to that of the theoretical model, with the variance reaching a stable plateau with minimal fluctuations [Fig. \ref{fig:STEP}(c)].
The ability to steer the system back and forth between two equilibrium configurations, by acting on the interaction strength $g(t)$, foreshadows the possibility of devising quantum BEC engines, which will be addressed in the next paragraphs.

{\it Optimal protocols.}--- The protocol of Fig. \ref{fig:STEP} allowed us to drive the system to a new equilibrium in a time corresponding to just a few trap periods. However, one can do even better by deriving an optimal protocol that minimizes the duration of the transition.
In order to do so, we adopt the method illustrated in our earlier works \cite{hardel_shortcuts_2024,rosales-cabara_optimal_2020}. We first change independent variable from the time $t$ to the variance $s$. Then, we construct the cost functional to be minimized, which is a function of the  classical stiffness $\bar\kappa(s)$ and its derivative $\bar{\kappa}^{\prime}(s)$~\footnote{For simplicity, we use the same symbol for a function of time such as $\bar\kappa(t)$ and the corresponding function of the variance $\bar\kappa(s)=\bar\kappa(t(s))$.}

\begin{equation} \label{eq:cost}
    J[\bar{\kappa},\bar{\kappa}^{\prime}]=\int^{s_{f}}_{s_{i}} \left( \frac{1}{2}\frac{\gamma}{D\gamma -s\bar{\kappa}(s)} + \lambda f(s,\bar{\kappa},\bar{\kappa}^{\prime}) + \mu   
    \abs{\bar{\kappa}^{\prime}}^{2}\right)\mathrm{d}s,
\end{equation}
where the first term represents the total duration of the protocol 
$\Delta t = \frac{\gamma}{2} \int^{s_{f}}_{s_{i}} ds / [D\gamma-\bar{\kappa}(s)s]$,
obtained from Eq. \eqref{eq:variance}. The second term is an arbitrary cost to be defined later, and the last term limits the gradient of the stiffness. $\lambda$ and $\mu$ are Lagrange weights that modulate the relative importance of each term. Their choice determines, for instance, the duration $\Delta t$ of the protocol. 

The choice of the cost function $f(s,\bar{\kappa},\bar{\kappa}^{\prime})$ is dictated by the quantity that the user wishes to optimize alongside the duration of the protocol. Following our earlier work \cite{hardel_shortcuts_2024}, we choose to minimize the phase of the wave function $\alpha(t)=\frac{m}{2\hbar}\frac{\dot \sigma}{\sigma}$ (in absolute value)~\footnote{Other choices were explored in \cite{hardel_shortcuts_2024}.}. This choice yields the cost function $f(s,\bar{\kappa},\bar{\kappa}^{\prime})=\frac{m^{2}}{8\gamma\hbar^{2}}\,(D\gamma -s\bar{\kappa})/s^{2}$, and the corresponding Euler-Lagrange equation
\begin{equation}\label{eq:alpha_opt}
    2\mu \frac{{\rm d}^2 \bar\kappa}{{\rm d}s^2} = \frac{\gamma s}{{\left(D\gamma - s\bar{\kappa} \right)}^{2}} - \lambda \frac{m^{2}}{8\gamma\hbar^{2}s}
\end{equation}
(see Appendix \ref{app:functional} for details).
Solving Eq. \eqref{eq:alpha_opt} with boundary conditions at equilibrium $\bar{\kappa}_{i,f}= D\gamma/s_{i,f}$ yields a protocol with optimal trade-off between total duration, phase, and gradient of the stiffness. Such a trade-off can be modulated through the choice of the Lagrange weights $\lambda$ and $\mu$.
We note that, for an adiabatic process, the phase is $\alpha(t) \approx 0$, because all quantities vary infinitely slowly. Hence, our choice to minimize $|\alpha|$ is tantamount to constructing protocols that are ``adiabatically optimal", i.e., they are as close as possible to an adiabatic process compatibly with a finite duration.

\begin{figure}
\includegraphics[width=0.9\linewidth]{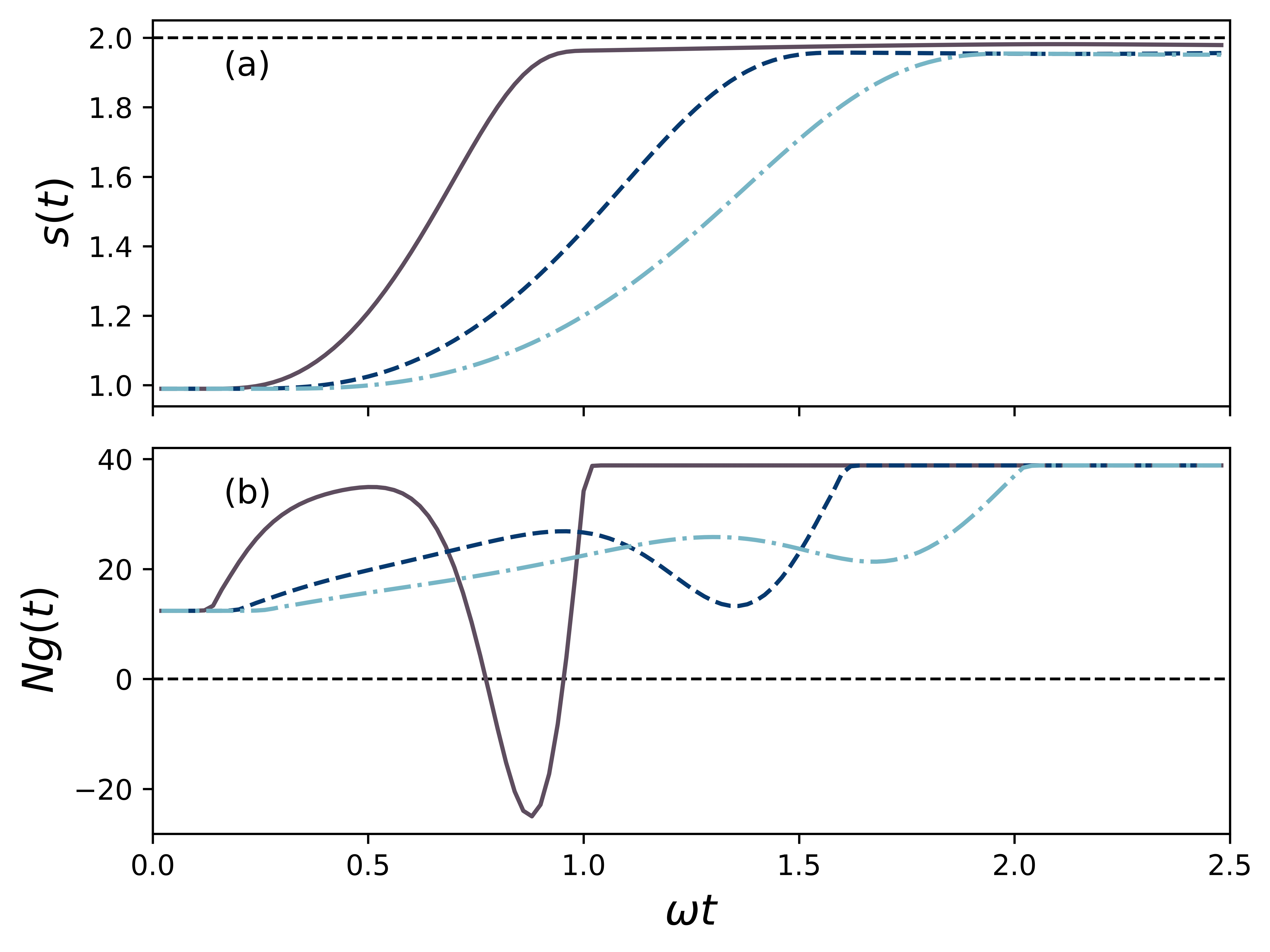}
    \caption{Optimal protocol with $\lambda=8$ and three values of the weight $\mu=0.1$ ($\omega \Delta t = 1.06$, solid purple line), $\mu=0.5$ ($\omega \Delta t = 1.68$, dashed dark-blue line), and $\mu=1$ ($\omega \Delta t = 2.11$, dot-dashed light-blue line), where $\Delta t$ is the duration of the protocol. The top panel (a) displyas the variance computed from the GPE using the optimal protocol $g(t)$ shown in panel (b).
    The runs are continued after the end of the protocols, keeping $g$ constant and equal to its final value, in order to highlight their stability. The variance $s$ is expressed in units of $L^2_{ho} = \hbar/(m\omega)$ and $g$ in units of $\hbar \omega L_{ho}$.}
\label{fig:optimal}
\end{figure}

Figure \ref{fig:optimal} shows three optimal $g$-protocols that transfer the variance from $s_i=1$ to $s_f=2$, for $\lambda = 8$ and three values of $\mu$, corresponding to different durations. The evolution of the variance in panel (a) has been obtained by direct numerical solution of the GPE with the optimal protocol $g(t)$ shown in panel (b). In order to highlight the remarkable stability of the solutions, the simulations have been extended beyond the end of the protocol, keeping $g(t)$ constant and equal to its final value. Such optimal protocols lead to a transition time $\Delta t$ that is significantly shorter than one oscillation period of the trap potential, $T=2\pi/\omega$, whereas, for the simple step of Fig. \ref{fig:STEP}, we had $\Delta t \approx 2.5 \,T$. Only for the shortest protocol does $g(t)$ take some negative values, which would be difficult, but not impossible, to realize in practice.

{\it Optimal Feshbach engines.}--- Here, we make full use of Eq. \eqref{eq:bridge} to devise efficient BEC cycles. This equation enables one to convert the BEC dynamics, governed by the GPE with varying stiffness $\kappa(t)$ and/or interaction strength $g(t)$, into a classical stochastic process where only the stiffness $\bar\kappa(t)$ is time-dependent.

In the simulation results, time is expressed in terms of the reference frequency $\omega_0/2\pi = 17.5 \,\rm Hz$ ($\omega_0^{-1} \approx 9.1 \,\rm ms$), the variance in units of $L^2_{ho} = \hbar/(m\omega_0)$, the stiffness in units of $m \omega_0^2$, and $g$ in units of $\hbar \omega_0 L_{ho}$. For instance, 
for lithium atoms $^{7}\rm Li$ \cite{Simmons2023}, one has $m=1.17 \times 10^{-26}\,\rm kg$, yielding $L_{ho} \approx 3.6 \rm \, \mu m$.


\begin{figure}
    \includegraphics[width=0.95\linewidth]{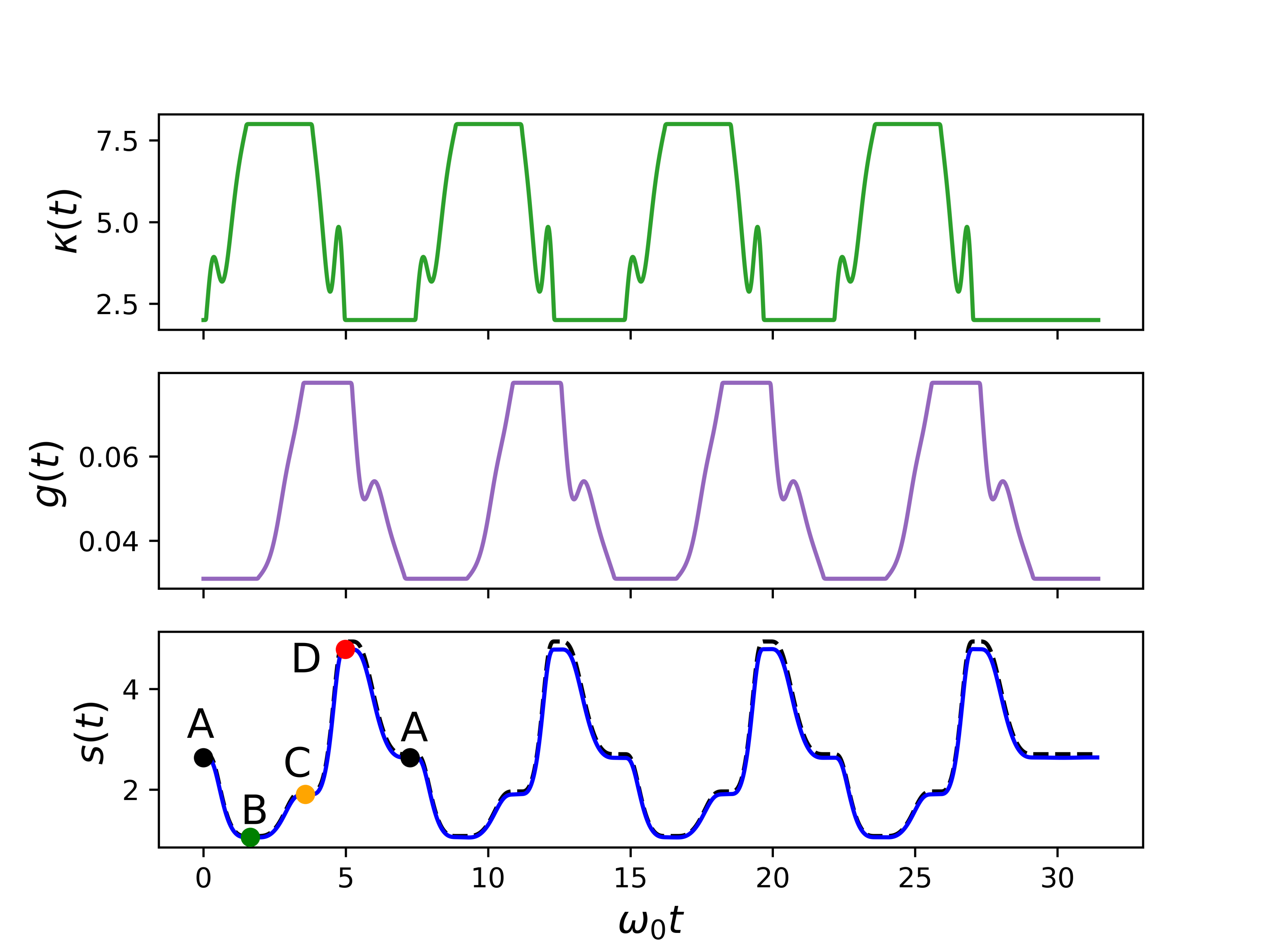}
    \caption{Four cycles of an optimal quantum engine for a BEC consisting of $N = 2000$ atoms. Each cycle alternates between $\kappa$-transitions and $g$-transitions. The top and middle panels show the evolutions of $\kappa(t)$ and $g(t)$. The bottom panel shows the variance expected from the theoretical protocols (dashed black line) and computed from simulations of the GPE (blue solid line). 
    The full ABCDA cycle is a sequence of 4 strokes: $\kappa_i \rightarrow \kappa_f$ at constant $g_i$ (AB); $g_i \rightarrow g_f$ at constant $\kappa_f$ (BC); $\kappa_f \rightarrow \kappa_i$ at constant $g_f$ (CD); $g_f \rightarrow g_i$ at constant $\kappa_i$ (DA). Here, $\kappa_i = 2$, $\kappa_f = 8$ (hence $\omega_i = \sqrt{2}$, $\omega_f = \sqrt{8}$), $g_i = 0.031$, and $g_f = 0.0775$, expressed in the normalized units defined in the main text.
   The initial and final variances are computed from Eq. \eqref{eq:bridge} at steady state.
   The Lagrange weights are: $\lambda = 0.01$ (for all strokes) and $\mu_{AB}=1$, $\mu_{BC}=4$, $\mu_{CD} = \mu_{DA} = 0.5$. 
}
  \label{fig:cycle-4}
\end{figure}

\begin{figure}
    \includegraphics[width=0.95\linewidth]{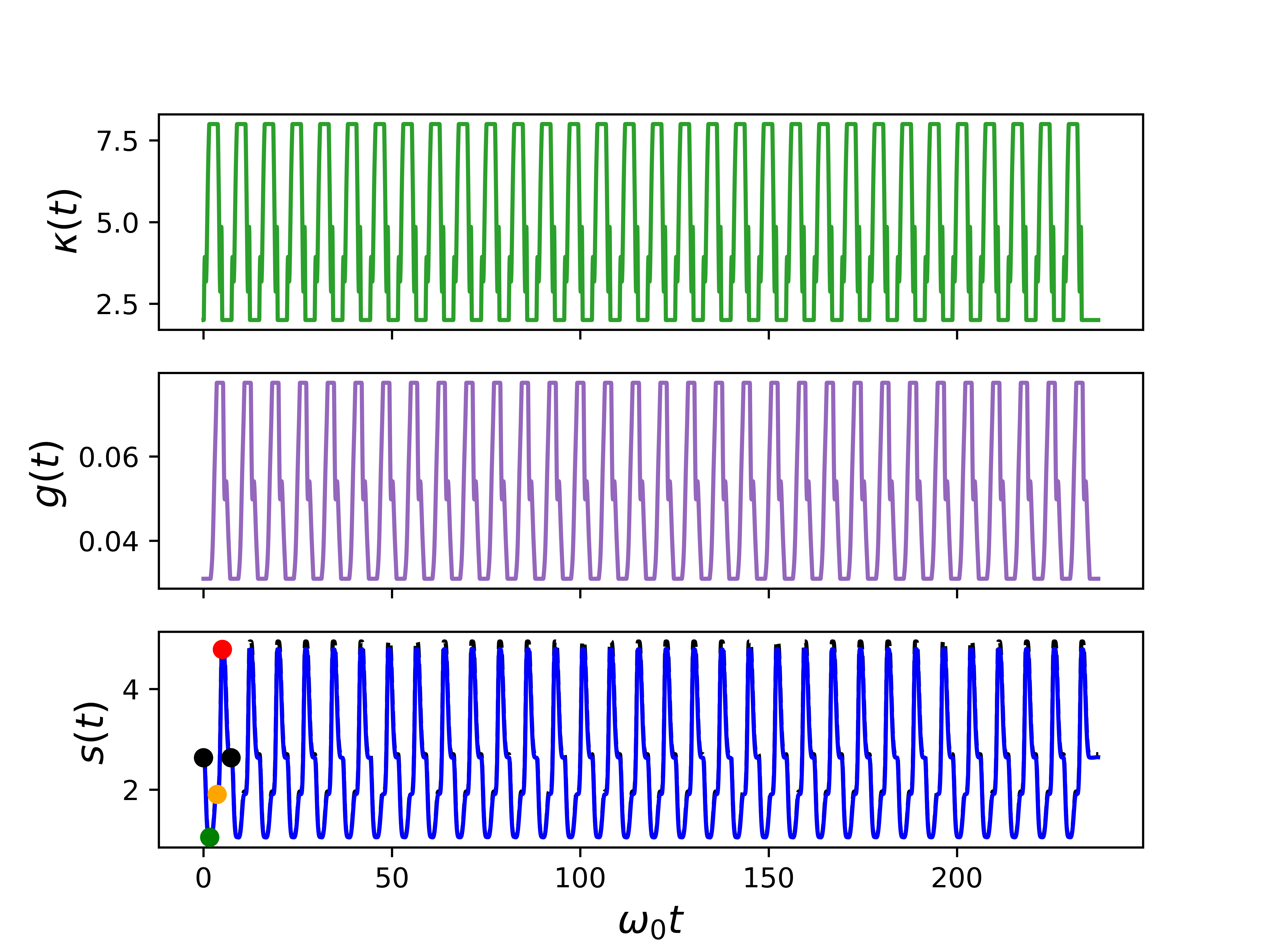}
    \caption{Same as Fig. \ref{fig:cycle-4}, for 32 repetitions of the cycle.}
\label{fig:cycle-32}
\end{figure}

In Fig. \ref{fig:cycle-4}, we consider an engine whose unit cycle ABCDA is made of 4 strokes that each minimize the phase $\alpha$ of the BEC wave function, thus making the overall cycle as close as possible to an adiabatic one, but with finite duration. During each stroke, either $\kappa $ is modulated at constant $g$ (steps AB and CD), or $g$ is varied at constant $\kappa$ (steps BC and DA). Such an engine was recently realized experimentally using a BEC of $^7\rm Li$ atoms \cite{Simmons2023}, albeit for an isotropic BEC and not a cigar-shaped one as considered here. We ran this quantum engine for a total of 4 cycles to highlight its remarkable stability. 
The bottom panel of Fig. \ref{fig:cycle-4} shows the evolution of $s(t)$ deduced from our optimization procedure (dashed line), together with the variance $\langle x^2 \rangle$ obtained from the numerical solution of the GPE, showing excellent agreement. The GPE is run some time after the end of the last cycle, with constant $g$ and $\kappa$, in order to highlight the robustness of the final state. 
In the Appendix \ref{app:reverse}, we show the reverse cycle ADCBA, which is just as stable as the forward one. The observed reversibility is a further signature that the cycle mimics an adiabatic one, but with finite duration.
Finally, in Fig. \ref{fig:cycle-32}, the same cycle is repeated 32 times, without any noticeable loss of stability.

{\it Power and efficiency.}---
The quantum engine studied above differs from ordinary thermal engines \cite{Watson2025,Campbell2026} inasmuch as it does not use heat to produce work---instead, it converts one type of energy into another. In our case, it converts the magnetic energy needed to modulate $g$ through the Feshbach resonance \cite{roati_k_2007,chin_feshbach_2010} into mechanical work.
Two important quantities to assess the quality of a cycle are its efficiency $\eta$ and the total delivered power $P$ per cycle, defined respectively as \cite{li_efficient_2018}
\[
\eta=-\frac{W_{AB}+W_{CD}}{W_{BC}}, \quad P=-\frac{W_{AB}+W_{CD}}{\tau_{\rm cycle}},
\]
where $\tau_{\rm cycle}$ is the total cycle time and  $W_{IJ}$ is the work done on the BEC during stroke $IJ$ of the cycle. If $I$ and $J$ are stationary states, then $W_{IJ}$ is equal to the energy difference $\Delta_{IJ} \equiv E_{J} - E_{I}$, where $E_{I,J}$ are steady-state energies computed from the GPE. By convention, $W>0$ when the system gains energy. Note that $P$ is an extensive quantity that depends on the number $N$ of atoms. If the target state is not stationary, then some irreversible work appears:  $W_{IJ} = \Delta_{IJ} +W_{\rm irr}$.
 We call an engine ``stable" when $W_{\rm irr} \approx 0$ after many cycles, and ``unstable" when $W_{\rm irr} \neq 0$.

\begin{figure}
    \includegraphics[width=0.9\linewidth]{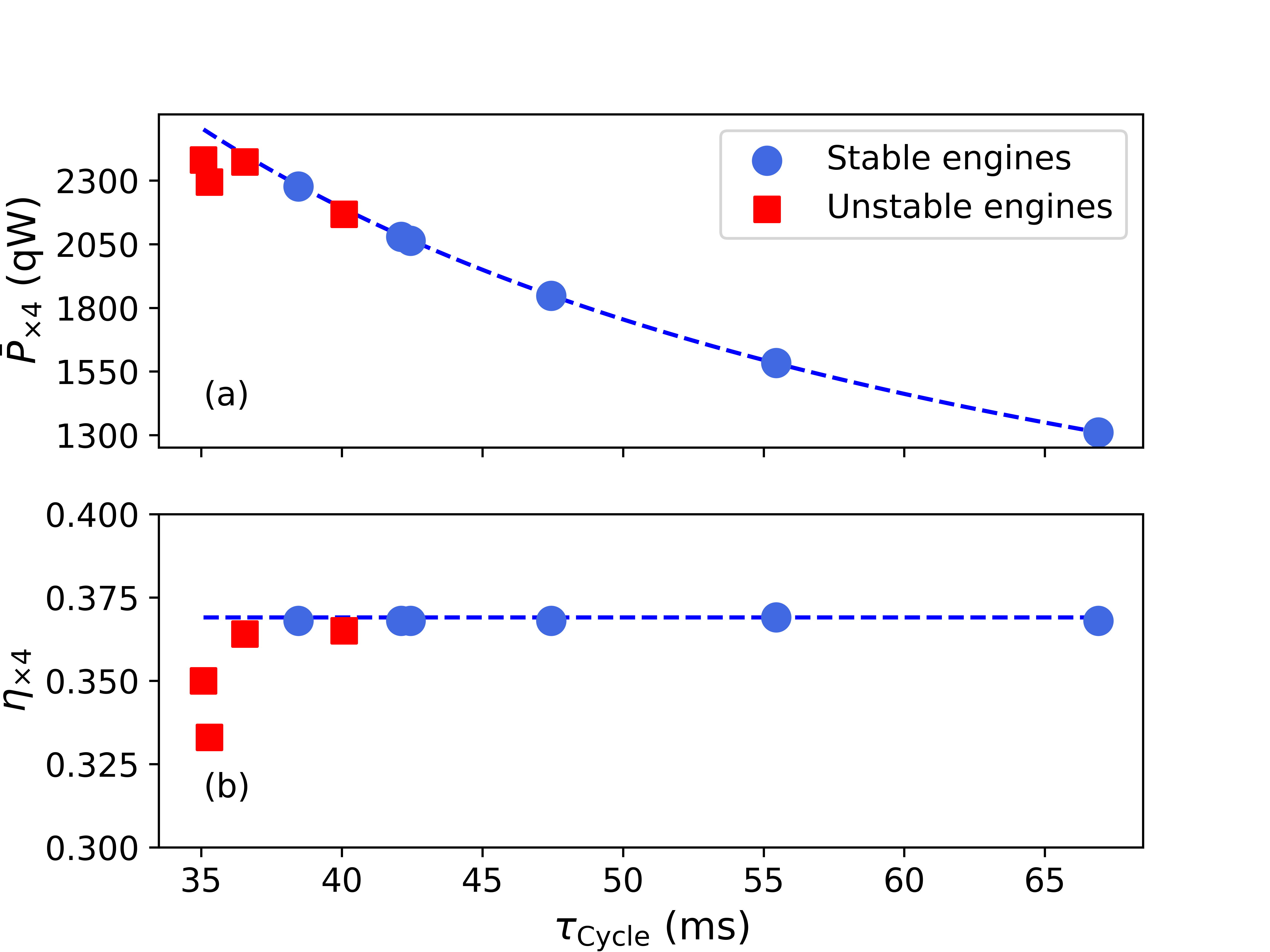}
    \caption{Total power (a) and efficiency (b), both averaged over 4 cycles, versus cycle time. The blue circles refer to stable cycles (irreversible work $W_{\rm irr} \ll \Delta_{IJ}$) and the red squares to unstable cycles. Power is measured in quectowatts ($\rm 1\, qW = 10^{-30}\, W$). The dashed lines show the ideal results for $W_{\rm irr} = 0$.}
    \label{fig:power}
\end{figure}

The efficiency and power for several quantum engines are plotted in Fig. \ref{fig:power} as a function of the cycle time. Power increases at shorter cycle times, as expected. Below a certain critical value of $\tau_{\rm cycle}^{\displaystyle *}$, some irreversible work appears and the cycle loses both power and efficiency. It can be checked that  $\tau_{\rm cycle}^{\displaystyle *} \approx 2\pi/\omega_i$, which is the longest period of the trap harmonic potential.
The extracted power is considerably larger than that reported in recent experiments \cite{Simmons2023}, while the cycle time is shorter. Despite differences in geometry (isotropic vs. cigar-shaped) and regime (Thomas-Fermi vs. diluted), these results suggest that our optimal control approach could significantly enhance the output power.

{\it Conclusions.}---
In this work, we have developed a comprehensive strategy to obtain optimally efficient quantum cycles for an ``engine" based on a trapped BEC. The cycle is based on alternate strokes that modify either the trap stiffness at constant interaction strength (tuned via the Feshbach resonance) or the interaction strength at constant stiffness. Such engines, which convert magnetic energy into mechanical energy and viceversa, have recently been realized experimentally \cite{Simmons2023}. A variational approach enabled us to optimize each stroke of the cycle by minimizing a user-defined cost functional.
The same technique may apply to other ultracold atomic gases, such as Fermi-Bose mixtures at the BEC-BCS crossover \cite{Koch2023,Menon2025}.

Moreover, our work illustrates a general procedure to construct optimal protocols for systems governed by a nonlinear Schr\"odinger equation (NLSE): (i) Posit a Gaussian ansatz for the wave function, (ii) obtain an approximate solution of the NLSE through a variational method, (iii) derive the classical stochastic analog (Ornstein-Uhlenbeck) using Nelson's method, (iv) define an appropriate cost functional to be minimized, and finally (v) solve the corresponding Euler-Lagrange equation to obtain the optimal protocol. This versatile approach opens the door to future applications in fields as varied as nonlinear optics, quantum fluids, and quantum plasmas.

\acknowledgments{This work of the Interdisciplinary Thematic Institute QMat, as part of the ITI 2021--2028 program of the University of Strasbourg, CNRS and Inserm, was supported by IdEx Unistra (ANR 10 IDEX 0002), and by the SFRI STRAT’US project (ANR 20 SFRI 0012) and EUR QMAT ANR-17-EURE-0024 under the framework of the French Investments for the Future Program. 
}


\appendix

\section{Variational solution of the GPE}\label{app:variational}

In order to obtain an approximate solution of Eq. \eqref{eq:GP}, we write the wave function in the Madelung representation  \cite{Haas2009,Hurst2016}: $\psi(x,t)=\sqrt{P(x,t)}e^{iS(x,t)}$, where the probability density $P$ and the phase $S$ are real functions. Then, the GPE is equivalent to the  Lagrangian density
\begin{equation} \label{eq:Ldensity}
    \mathcal{L}[P,S]=-\hbar P\partial_{t}S -\frac{\hbar^{2}}{2m} \left[P{\left(\partial_{x}S \right)}^{2} + \frac{(\partial_{x}P)^{2}}{4P}\right]- U P,
\end{equation}
where $U(x,t)=\frac{1}{2}[m\omega^{2}x^{2} + NgP(x,t)]$. The Euler-Lagrange equations for $\mathcal{L}$ yield the continuity equation
\begin{equation}
    \partial_{t} P +\frac{\hbar}{m}\partial_{x}\left(P\partial_{x}S\right) = 0,
    \label{eq:continuity}
\end{equation}
and the quantum Hamilton-Jacobi equation
\begin{equation}
   \hbar\partial_{t}S + \frac{\hbar^{2}}{2m}\left( \partial_{x}S \right)^{2} +\frac{1}{2}m\omega^{2}x^{2} +gP - \frac{\hbar^{2}}{2m}\frac{\partial_{x}^{2} \sqrt{P}}{\sqrt{P}}=0,
    \label{eq:HJ}
\end{equation}
where $\hbar \partial_x S/m$ is the fluid velocity. Equations \eqref{eq:continuity} and \eqref{eq:HJ} are equivalent to the GPE expressed in the Madelung representation  \cite{Haas2009,Hurst2016}.

We now assume that the density is Gaussian and the phase is quadratic in $x$
\[
P(x,t)=\frac{1}{\sqrt{2\pi}\sigma(t)}\, e^{-x^2/2\sigma^2(t)}, \quad S(x,t)=\alpha(t) x^{2}.
\]
From the continuity equation for the density $P$, it can be easily deduced that $\alpha(t)=\frac{m}{2\hbar}\frac{\dot \sigma}{\sigma}$.
Injecting the expressions for $P$ and $S$ into the Lagrangian density \eqref{eq:Ldensity} and integrating over the spatial co-ordinate, one obtains the Lagrangian $L= \int_{-\infty}^{\infty} \mathcal{L}\,dx$ as
\begin{equation} \label{eq:Lagrangian}
L[\sigma,\ddot\sigma]=-\frac{m}{2}\ddot\sigma \sigma - \frac{\hbar^{2}}{8m\sigma^{2}} - \frac{1}{2}m\omega^{2}\sigma^{2}- \frac{Ng}{4\sqrt{\pi}\sigma} .
\end{equation}
The corresponding Euler-Lagrange equation yields the equation of motion for the mean-square displacement $\sigma(t)$
\begin{equation} \label{eq:Ermakov}
    \ddot \sigma + \omega^{2} \sigma= \frac{\hbar^{2}}{4m^{2}\sigma^{3}} + \frac{Ng}{4\sqrt{\pi}m\sigma^{2}},
\end{equation}
which is a modified Ermakov equation \cite{Lewis1967}. Plugging the solution of Eq. \eqref{eq:Ermakov} into the Gaussian ansatz for $\psi(x,t)$ provides the Gaussian wave function that best approximates the exact solution of the GPE.

Unlike similar STA approaches \cite{Lewis1967,Schaff2011,Mishra2025}, the Ermakov equation \eqref{eq:Ermakov} explicitly contains the interaction strength $g$. Hence, the approximate Gaussian solution naturally adapts itself to the different regimes of weak or strong interactions and remains reasonably accurate over a wide range of parameters, as was proven for the case of an interacting Coulomb gas \cite{Haas2009}.


\section{Quantum-classical analogy}\label{app:bridge}

Equating the  stiffness of the classical Ornstein-Uhlenbeck process $\bar{\kappa}(t)$ to the stiffness of the quantum Nelson equation \eqref{eq:OUquantum}, we get
\[
   -\frac{\bar{\kappa}}{\gamma} =  \frac{\hbar}{m} \left(2\alpha-\frac{1}{2\sigma^{2}} \right),
\]
whose time derivative yields
\[
\dot{\bar{\kappa}}=-\frac{\hbar\gamma}{m} \left(\frac{m}{\hbar}\frac{\ddot \sigma \sigma-{\dot\sigma}^{2}}{\sigma^{2}}+\frac{\dot \sigma}{\sigma^{3}} \right) .
\]
Solving for $\ddot \sigma $ and substituting into Eq. \eqref{eq:Ermakov}, we obtain
\[
\kappa -\frac{m\dot{\bar{\kappa}}}{\gamma} -\hbar\frac{\dot{\sigma}}{\sigma^{3}} + \frac{m\dot{\sigma}^{2}}{\sigma^{2}} = \frac{\hbar^{2}}{4m\sigma^{4}} + \frac{Ng}{4\sqrt{\pi}\sigma^{3}},
\]
where we used $\kappa = m\omega^2$.
Remembering that $\sigma^2 = s$ and using Eq. \eqref{eq:variance} to express $\dot \sigma$ as
\[
\dot\sigma=\frac{1}{\gamma\sigma}(D\gamma-\bar{\kappa}\sigma^{2}),
\]
one finally obtains Eq. \eqref{eq:bridge}.


\section{Cost functional}\label{app:functional}

We want to minimize, in absolute value, the phase $\alpha(t)$ of the wave function \cite{hardel_shortcuts_2024}. The corresponding cost functional is then
\begin{equation*}
    \begin{split}
        F[\bar{\kappa},\bar{\kappa}^{\prime}] & =\int^{t_{f}}_{t_{i}}\alpha^{2}(t)\,\mathrm{d}t \\
        & = \int^{t_{f}}_{t_{i}}\left(\frac{m}{4\hbar}\frac{\dot{s}(t)}{s(t)}\right)^{2}\mathrm{d}t \\
        & = \int_{s_{i}}^{s_{f}}\left(\frac{m}{4\hbar s}\frac{2\left(D\gamma -s\bar{\kappa}\right)}{\gamma}\right)^{2} \, \frac{\gamma }{2\left(D\gamma -s\bar{\kappa}\right)}\,\mathrm{d}s \\
        & = \frac{m^{2}}{8\gamma\hbar^{2}}\int^{s_{f}}_{s_{i}}\frac{D\gamma -s\bar{\kappa}}{s^{2}}\,\mathrm{d}s,
    \end{split}
\end{equation*}
where we have used Eq. \eqref{eq:variance}.
From this, we immediately obtain the function
\begin{equation}
    f(s,\bar{\kappa},\bar{\kappa}^{\prime}) = \frac{m^{2}}{8\gamma\hbar^{2}}\frac{D\gamma -s\bar{\kappa}}{s^{2}}, 
    \label{eq:f}
\end{equation}
which can be inserted into the total cost functional $J[\bar{\kappa},\bar{\kappa}^{\prime}]$ defined in  Eq. \eqref{eq:cost}.

Such cost functional can be written as
\[
    J[\bar{\kappa},\bar{\kappa}^{\prime}] = \int^{s_{f}}_{s_{i}} L(s,\bar{\kappa},\bar{\kappa}^{\prime})\,\mathrm{d}s,
\]
where the Lagrangian is
 \[
    L[s,\bar{\kappa},\bar{\kappa}^{\prime}]=\frac{1}{2}\frac{\gamma}{D\gamma -s\bar{\kappa}(s)} + \lambda f(s,\bar{\kappa},\bar{\kappa}^{\prime}) + \mu \abs{\bar{\kappa}^{\prime}}^{2}.
 \]
The associated Euler-Lagrange equation 
$$
\partialderivative{L}{\bar{\kappa}}-\derivative{}{s}\partialderivative{L}{\bar{\kappa}^{\prime}}=0
$$ 
yields 
\begin{equation}
    2\mu\, \frac{{\rm d^2} \bar{\kappa}}{{\rm d} s^2} = \frac{\gamma s}{{\left(D\gamma - s\bar{\kappa} \right)}^{2}} + \lambda\partialderivative{f}{\bar{\kappa}}-\lambda \derivative{}{s}\partialderivative{f}{\bar{\kappa}^{\prime}}.
    \label{eq:kprimeprime}
\end{equation}
Injecting the expression \eqref{eq:f} into Eq. \eqref{eq:kprimeprime}, we finally obtain Eq. \eqref{eq:alpha_opt}.
Equation \eqref{eq:kprimeprime} can be solved numerically for $\bar \kappa(s)$ as a boundary-value problem on the interval $[s_{i},s_{f}]$, with boundary conditions $\bar{\kappa}(s_{i,f})= D\gamma/s_{i,f}$ corresponding to steady-state solutions of Eq. \eqref{eq:variance}.

From the solution $\bar \kappa (s)$ expressed in terms of the variance $s$, one obtains $\bar \kappa$ as a function of time $\bar \kappa (t)= \bar \kappa (s(t))$ by solving Eq. \eqref{eq:variance}. The corresponding control parameters $\kappa(t)$ or $g(t)$ then follow directly from Eq. \eqref{eq:bridge}.


\section{Reverse cycle}\label{app:reverse}
\begin{figure}
    \includegraphics[width=0.95\linewidth]{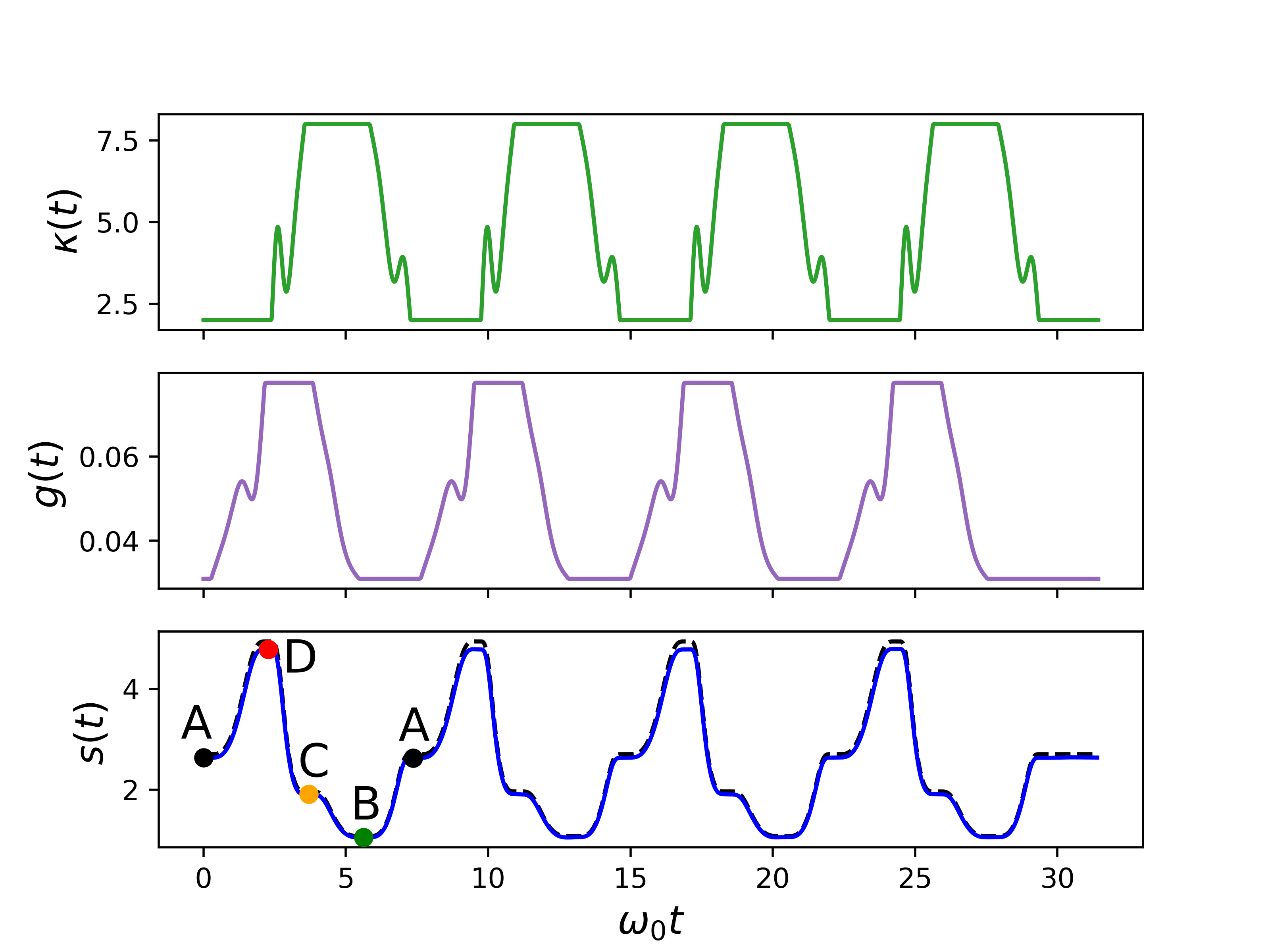}
    \caption{Same as Fig. \ref{fig:cycle-4} for the reverse cycle ADCBA.}
  \label{fig:cycle-1-rev}
\end{figure}

In Fig. \ref{fig:cycle-1-rev}, we show the reverse cycle ADCBA corresponding to the direct cycle displayed in Fig. \ref{fig:cycle-4}. The reverse cycle is as stable as the direct one.

\newpage
\bibliography{Rapport-Nelson-BEC}

@article{Menon2025,
doi = {10.1088/2058-9565/ae01d3},
url = {https://doi.org/10.1088/2058-9565/ae01d3},
year = {2025},
month = {sep},
publisher = {IOP Publishing},
volume = {10},
number = {4},
pages = {045039},
author = {Menon, Keerthy and Busch, Thomas and Fogarty, Thomás},
title = {Leveraging quantum statistics to enhance heat engines},
journal = {Quantum Science and Technology},
}

@article{Mishra2025,
doi = {10.1088/1367-2630/adb905},
url = {https://doi.org/10.1088/1367-2630/adb905},
year = {2025},
month = {mar},
publisher = {IOP Publishing},
volume = {27},
number = {3},
pages = {033009},
author = {Mishra, Chinmayee and Busch, Thomas and Fogarty, Thomás},
title = {Shortcuts to adiabaticity in anisotropic {Bose-Einstein} condensates},
journal = {New Journal of Physics},
}

@article{Brandner2015,
doi = {10.1088/1367-2630/17/6/065006},
url = {https://doi.org/10.1088/1367-2630/17/6/065006},
year = {2015},
month = {jun},
publisher = {IOP Publishing},
volume = {17},
number = {6},
pages = {065006},
author = {Brandner, Kay and Bauer, Michael and Schmid, Michael T and Seifert, Udo},
title = {Coherence-enhanced efficiency of feedback-driven quantum engines},
journal = {New Journal of Physics},
}

@Article{Beau2016,
AUTHOR = {Beau, Mathieu and Jaramillo, Juan and Del Campo, Adolfo},
TITLE = {Scaling-Up Quantum Heat Engines Efficiently via Shortcuts to Adiabaticity},
JOURNAL = {Entropy},
VOLUME = {18},
YEAR = {2016},
NUMBER = {5},
ARTICLE-NUMBER = {168},
URL = {https://www.mdpi.com/1099-4300/18/5/168},
ISSN = {1099-4300},
DOI = {10.3390/e18050168}
}

@article{Schaff2011,
doi = {10.1209/0295-5075/93/23001},
url = {https://doi.org/10.1209/0295-5075/93/23001},
year = {2011},
month = {jan},
publisher = {},
volume = {93},
number = {2},
pages = {23001},
author = {Schaff, J.-F. and Song, X.-L. and Capuzzi, P. and Vignolo, P. and Labeyrie, G.},
title = {Shortcut to adiabaticity for an interacting {Bose-Einstein} condensate},
journal = {Europhysics Letters},
}

@article{FengWu2006,
title = {Quantum degeneracy effect on performance of irreversible Otto cycle with ideal Bose gas},
journal = {Energy Conversion and Management},
volume = {47},
number = {18},
pages = {3008-3018},
year = {2006},
issn = {0196-8904},
doi = {https://doi.org/10.1016/j.enconman.2006.03.011},
url = {https://www.sciencedirect.com/science/article/pii/S0196890406001038},
author = {Feng Wu and Lingen Chen and Fengrui Sun and Chih Wu and Fangzhong Guo and Qing Li},
}

@article{Estrada2024,
  title = {Quantum engines with interacting {Bose-Einstein} condensates},
  author = {Estrada, Juli\'an Amette and Mayo, Franco and Roncaglia, Augusto J. and Mininni, Pablo D.},
  journal = {Phys. Rev. A},
  volume = {109},
  issue = {1},
  pages = {012202},
  numpages = {9},
  year = {2024},
  month = {Jan},
  publisher = {American Physical Society},
  doi = {10.1103/PhysRevA.109.012202},
  url = {https://link.aps.org/doi/10.1103/PhysRevA.109.012202}
}

@article{Hurst2016,
  title = {High-harmonic generation in a quantum electron gas trapped in a nonparabolic and anisotropic well},
  author = {Hurst, J\'er\^ome and L\'ev\^eque-Simon, K\'evin and Hervieux, Paul-Antoine and Manfredi, Giovanni and Haas, Fernando},
  journal = {Phys. Rev. B},
  volume = {93},
  issue = {20},
  pages = {205402},
  numpages = {9},
  year = {2016},
  month = {May},
  publisher = {American Physical Society},
  doi = {10.1103/PhysRevB.93.205402},
  url = {https://link.aps.org/doi/10.1103/PhysRevB.93.205402}
}

@article{Haas2009,
  title = {Breather mode in the many-electron dynamics of semiconductor quantum wells},
  author = {Haas, F. and Manfredi, G. and Shukla, P. K. and Hervieux, P.-A.},
  journal = {Phys. Rev. B},
  volume = {80},
  issue = {7},
  pages = {073301},
  numpages = {4},
  year = {2009},
  month = {Aug},
  publisher = {American Physical Society},
  doi = {10.1103/PhysRevB.80.073301},
  url = {https://link.aps.org/doi/10.1103/PhysRevB.80.073301}
}

@Article{Watson2025,
	title={{Quantum many-body thermal machines enabled by atom-atom correlations}},
	author={Raymon S. Watson and Karen V. Kheruntsyan},
	journal={SciPost Phys.},
	volume={18},
	pages={190},
	year={2025},
	publisher={SciPost},
	doi={10.21468/SciPostPhys.18.6.190},
	url={https://scipost.org/10.21468/SciPostPhys.18.6.190},
}

@article{Nautiyal2024,
doi = {10.1088/1367-2630/ad57e5},
url = {https://doi.org/10.1088/1367-2630/ad57e5},
year = {2024},
month = {jun},
publisher = {IOP Publishing},
volume = {26},
number = {6},
pages = {063033},
author = {Nautiyal, V V and Watson, R S and Kheruntsyan, K V},
title = {A finite-time quantum {Otto} engine with tunnel coupled one-dimensional {Bose} gases},
journal = {New Journal of Physics},
}

@article{Campbell2026,
doi = {10.1088/2058-9565/ae1e27},
url = {https://doi.org/10.1088/2058-9565/ae1e27},
year = {2026},
month = {jan},
publisher = {IOP Publishing},
volume = {11},
number = {1},
pages = {012501},
author = {Campbell, Steve and D’Amico, Irene and Ciampini, Mario A and Anders, Janet and Ares, Natalia and Artini, Simone and Auffèves, Alexia and Bassman Oftelie, Lindsay and Bettmann, Laetitia P and Bonança, Marcus V S and Busch, Thomas and Campisi, Michele and Cavalcante, Moallison F and Correa, Luis A and Cuestas, Eloisa and Dag, Ceren B and Dago, Salambô and Deffner, Sebastian and Del Campo, Adolfo and Deutschmann-Olek, Andreas and Donadi, Sandro and Doucet, Emery and Elouard, Cyril and Ensslin, Klaus and Erker, Paul and Fabbri, Nicole and Fedele, Federico and Fiusa, Guilherme and Fogarty, Thomás and Folk, Joshua and Guarnieri, Giacomo and Hegde, Abhaya S and Hernández-Gómez, Santiago and Hu, Chang-Kang and Iemini, Fernando and Karimi, Bayan and Kiesel, Nikolai and Landi, Gabriel T and Lasek, Aleksander and Lemziakov, Sergei and Lo Monaco, Gabriele and Lutz, Eric and Lvov, Dmitrii and Maillet, Olivier and Mehboudi, Mohammad and Mendonça, Taysa M and Miller, Harry J D and Mitchell, Andrew K and Mitchison, Mark T and Mukherjee, Victor and Paternostro, Mauro and Pekola, Jukka and Perarnau-Llobet, Martí and Poschinger, Ulrich and Rolandi, Alberto and Rosa, Dario and Sánchez, Rafael and Santos, Alan C and Sarthour, Roberto S and Sela, Eran and Solfanelli, Andrea and Souza, Alexandre M and Splettstoesser, Janine and Tan, Dian and Tesser, Ludovico and Van Vu, Tan and Widera, Artur and Yunger Halpern, Nicole and Zawadzki, Krissia},
title = {Roadmap on quantum thermodynamics},
journal = {Quantum Science and Technology},
}

@article{Koch2023,
  title={A quantum engine in the {BEC--BCS} crossover},
  author={Koch, Jennifer and Menon, Keerthy and Cuestas, Eloisa and Barbosa, Sian and Lutz, Eric and Fogarty, Thom{\'a}s and Busch, Thomas and Widera, Artur},
  journal={Nature},
  volume={621},
  number={7980},
  pages={723--727},
  year={2023},
  url = {https://doi.org/10.1038/s41586-023-06469-8},
  publisher={Nature Publishing Group UK London}
}

@article{Manfredi2008,
  title = {Fidelity Decay in Trapped {Bose-Einstein} Condensates},
  author = {Manfredi, G. and Hervieux, P.-A.},
  journal = {Phys. Rev. Lett.},
  volume = {100},
  issue = {5},
  pages = {050405},
  numpages = {4},
  year = {2008},
  month = {Feb},
  publisher = {American Physical Society},
  doi = {10.1103/PhysRevLett.100.050405},
  url = {https://link.aps.org/doi/10.1103/PhysRevLett.100.050405}
}

@Article{GueryOdelin2019,
  author    = {Gu\'ery-Odelin, D. and Ruschhaupt, A. and Kiely, A. and Torrontegui, E. and Mart\'{\i}nez-Garaot, S. and Muga, J. G.},
  journal   = {Rev. Mod. Phys.},
  title     = {Shortcuts to adiabaticity: Concepts, methods, and applications},
  year      = {2019},
  month     = {Oct},
  pages     = {045001},
  volume    = {91},
  doi       = {10.1103/RevModPhys.91.045001},
  issue     = {4},
  numpages  = {54},
  publisher = {American Physical Society},
  url       = {https://link.aps.org/doi/10.1103/RevModPhys.91.045001},
}

@article{Simmons2023,
  title = {Thermodynamic engine with a quantum degenerate working fluid},
  author = {Simmons, Ethan Q. and Sajjad, Roshan and Keithley, Kimberlee and Mas, Hector and Tanlimco, Jeremy L. and Nolasco-Martinez, Eber and Bai, Yifei and Fredrickson, Glenn H. and Weld, David M.},
  journal = {Phys. Rev. Res.},
  volume = {5},
  issue = {4},
  pages = {L042009},
  numpages = {8},
  year = {2023},
  month = {Oct},
  publisher = {American Physical Society},
  doi = {10.1103/PhysRevResearch.5.L042009},
  url = {https://link.aps.org/doi/10.1103/PhysRevResearch.5.L042009}
}

@article{Bacciagaluppi1999,
  title={Nelsonian mechanics revisited},
  author={Bacciagaluppi, Guido},
  journal={Foundations of Physics Letters},
  volume={12},
  number={1},
  pages={1--16},
  year={1999},
  publisher={Springer}
}

@article{Masuda2010,
    author = {Masuda, Shumpei and Nakamura, Katsuhiro},
    title = {Fast-forward of adiabatic dynamics in quantum mechanics},
    journal = {Proceedings of the Royal Society A: Mathematical, Physical and Engineering Sciences},
    volume = {466},
    number = {2116},
    pages = {1135-1154},
    year = {2009},
    month = {11},
    issn = {1364-5021},
    doi = {10.1098/rspa.2009.0446},
    url = {https://doi.org/10.1098/rspa.2009.0446},
}

@article{chin_feshbach_2010,
	title = {Feshbach resonances in ultracold gases},
	volume = {82},
	copyright = {http://link.aps.org/licenses/aps-default-license},
	issn = {0034-6861, 1539-0756},
	url = {https://link.aps.org/doi/10.1103/RevModPhys.82.1225},
	doi = {10.1103/RevModPhys.82.1225},
	language = {english},
	number = {2},
	urldate = {2025-05-07},
	journal = {Reviews of Modern Physics},
	author = {Chin, Cheng and Grimm, Rudolf and Julienne, Paul and Tiesinga, Eite},
	month = apr,
	year = {2010},
	pages = {1225--1286},
}

@article{nelson_derivation_1966,
	title = {Derivation of the {Schrödinger} {Equation} from {Newtonian} {Mechanics}},
	volume = {150},
	copyright = {http://link.aps.org/licenses/aps-default-license},
	issn = {0031-899X},
	url = {https://link.aps.org/doi/10.1103/PhysRev.150.1079},
	doi = {10.1103/PhysRev.150.1079},
	language = {english},
	number = {4},
	urldate = {2025-03-12},
	journal = {Physical Review},
	author = {Nelson, Edward},
	month = oct,
	year = {1966},
	pages = {1079--1085},
}

@article{castin_bose-einstein_2001,
	title = {Bose-{Einstein} condensates in atomic gases: simple theoretical results},
	copyright = {Assumed arXiv.org perpetual, non-exclusive license to distribute this article for submissions made before January 2004},
	shorttitle = {Bose-{Einstein} condensates in atomic gases},
	url = {https://arxiv.org/abs/cond-mat/0105058},
	doi = {10.48550/arxiv.con-mat/0105058},
	urldate = {2025-03-13},
	author = {Castin, Yvan},
	journal = {ArXiv},
	year = {2001},
}

@article{hardel_shortcuts_2024,
	title = {Shortcuts to adiabaticity in harmonic traps: {A} quantum-classical analog},
	volume = {110},
	issn = {2470-0045, 2470-0053},
	shorttitle = {Shortcuts to adiabaticity in harmonic traps},
	url = {https://link.aps.org/doi/10.1103/PhysRevE.110.054138},
	doi = {10.1103/PhysRevE.110.054138},
	language = {english},
	number = {5},
	urldate = {2025-03-13},
	journal = {Physical Review E},
	author = {Hardel, Vincent and Manfredi, Giovanni and Hervieux, Paul-Antoine and Goerlich, Rémi},
	month = nov,
	year = {2024},
	pages = {054138},
}

@article{fort_effect_2005,
	title = {Effect of {Optical} {Disorder} and {Single} {Defects} on the {Expansion} of a {Bose}-{Einstein} {Condensate} in a {One}-{Dimensional} {Waveguide}},
	volume = {95},
	issn = {0031-9007, 1079-7114},
	url = {https://link.aps.org/doi/10.1103/PhysRevLett.95.170410},
	doi = {10.1103/PhysRevLett.95.170410},
	language = {english},
	number = {17},
	urldate = {2025-03-14},
	journal = {Physical Review Letters},
	author = {Fort, C. and Fallani, L. and Guarrera, V. and Lye, J. E. and Modugno, M. and Wiersma, D. S. and Inguscio, M.},
	month = oct,
	year = {2005},
	pages = {170410},
}

@article{derrico_feshbach_2007,
	title = {Feshbach resonances in ultracold $^{\textrm{39}}${K}},
	volume = {9},
	issn = {1367-2630},
	url = {https://iopscience.iop.org/article/10.1088/1367-2630/9/7/223},
	doi = {10.1088/1367-2630/9/7/223},
	number = {7},
	urldate = {2025-07-04},
	journal = {New Journal of Physics},
	author = {D'Errico, Chiara and Zaccanti, Matteo and Fattori, Marco and Roati, Giacomo and Inguscio, Massimo and Modugno, Giovanni and Simoni, Andrea},
	month = jul,
	year = {2007},
	pages = {223--223},
}

@article{kagan_evolution_1997,
	title = {Evolution and {Global} {Collapse} of {Trapped} {Bose} {Condensates} under {Variations} of the {Scattering} {Length}},
	volume = {79},
	copyright = {http://link.aps.org/licenses/aps-default-license},
	issn = {0031-9007, 1079-7114},
	url = {https://link.aps.org/doi/10.1103/PhysRevLett.79.2604},
	doi = {10.1103/PhysRevLett.79.2604},
	language = {english},
	number = {14},
	urldate = {2025-04-01},
	journal = {Physical Review Letters},
	author = {Kagan, Yu. and Surkov, E. L. and Shlyapnikov, G. V.},
	month = oct,
	year = {1997},
	pages = {2604--2607},
}

@article{li_shortcut_2016,
	title = {Shortcut to adiabatic control of soliton matter waves by tunable interaction},
	volume = {6},
	issn = {2045-2322},
	url = {https://www.nature.com/articles/srep38258},
	doi = {10.1038/srep38258},
	language = {english},
	number = {1},
	urldate = {2025-04-04},
	journal = {Scientific Reports},
	author = {Li, Jing and Sun, Kun and Chen, Xi},
	month = dec,
	year = {2016},
	pages = {38258},
}

@article{li_efficient_2018,
	title = {An efficient nonlinear {Feshbach} engine},
	volume = {20},
	issn = {1367-2630},
	url = {https://iopscience.iop.org/article/10.1088/1367-2630/aa9cd8},
	doi = {10.1088/1367-2630/aa9cd8},
	number = {1},
	urldate = {2025-04-04},
	journal = {New Journal of Physics},
	author = {Li, Jing and Fogarty, Thomás and Campbell, Steve and Chen, Xi and Busch, Thomas},
	month = jan,
	year = {2018},
	pages = {015005},
}

@article{del_campo_frictionless_2011,
	title = {Frictionless quantum quenches in ultracold gases: {A} quantum-dynamical microscope},
	volume = {84},
	copyright = {http://link.aps.org/licenses/aps-default-license},
	issn = {1050-2947, 1094-1622},
	shorttitle = {Frictionless quantum quenches in ultracold gases},
	url = {https://link.aps.org/doi/10.1103/PhysRevA.84.031606},
	doi = {10.1103/PhysRevA.84.031606},
	language = {english},
	number = {3},
	urldate = {2025-04-04},
	journal = {Physical Review A},
	author = {Del Campo, A.},
	month = sep,
	year = {2011},
	pages = {031606},
}

@article{del_campo_shortcuts_2013,
	title = {Shortcuts to {Adiabaticity} by {Counterdiabatic} {Driving}},
	volume = {111},
	copyright = {http://link.aps.org/licenses/aps-default-license},
	issn = {0031-9007, 1079-7114},
	url = {https://link.aps.org/doi/10.1103/PhysRevLett.111.100502},
	doi = {10.1103/PhysRevLett.111.100502},
	language = {english},
	number = {10},
	urldate = {2025-04-04},
	journal = {Physical Review Letters},
	author = {Del Campo, Adolfo},
	month = sep,
	year = {2013},
	pages = {100502},
}

@article{rosales-cabara_optimal_2020,
	title = {Optimal protocols and universal time-energy bound in {Brownian} thermodynamics},
	volume = {2},
	issn = {2643-1564},
	url = {https://link.aps.org/doi/10.1103/PhysRevResearch.2.012012},
	doi = {10.1103/PhysRevResearch.2.012012},
	language = {english},
	number = {1},
	urldate = {2025-04-15},
	journal = {Physical Review Research},
	author = {Rosales-Cabara, Yoseline and Manfredi, Giovanni and Schnoering, Gabriel and Hervieux, Paul-Antoine and Mertz, Laurent and Genet, Cyriaque},
	month = jan,
	year = {2020},
	pages = {012012},
}

@article{roati_k_2007,
	title = {K39 {Bose}-{Einstein} {Condensate} with {Tunable} {Interactions}},
	volume = {99},
	copyright = {http://link.aps.org/licenses/aps-default-license},
	issn = {0031-9007, 1079-7114},
	url = {https://link.aps.org/doi/10.1103/PhysRevLett.99.010403},
	doi = {10.1103/PhysRevLett.99.010403},
	language = {english},
	number = {1},
	urldate = {2025-04-23},
	journal = {Physical Review Letters},
	author = {Roati, G. and Zaccanti, M. and D’Errico, C. and Catani, J. and Modugno, M. and Simoni, A. and Inguscio, M. and Modugno, G.},
	month = jul,
	year = {2007},
	pages = {010403},
}

@article{keller_feshbach_2020,
	title = {Feshbach engine in the {Thomas}-{Fermi} regime},
	volume = {2},
	issn = {2643-1564},
	url = {https://link.aps.org/doi/10.1103/PhysRevResearch.2.033335},
	doi = {10.1103/PhysRevResearch.2.033335},
	language = {english},
	number = {3},
	urldate = {2025-04-24},
	journal = {Physical Review Research},
	author = {Keller, Tim and Fogarty, Thomás and Li, Jing and Busch, Thomas},
	month = aug,
	year = {2020},
	pages = {033335},
}

@article{uhlenbeck_theory_1930,
	title = {On the {Theory} of the {Brownian} {Motion}},
	volume = {36},
	copyright = {http://link.aps.org/licenses/aps-default-license},
	issn = {0031-899X},
	url = {https://link.aps.org/doi/10.1103/PhysRev.36.823},
	doi = {10.1103/PhysRev.36.823},
	language = {english},
	number = {5},
	urldate = {2025-04-25},
	journal = {Physical Review},
	author = {Uhlenbeck, G. E. and Ornstein, L. S.},
	month = sep,
	year = {1930},
	pages = {823--841},
}

@incollection{torrontegui_shortcuts_2013,
	title = {Shortcuts to {Adiabaticity}},
	volume = {62},
	isbn = {978-0-12-408090-4},
	url = {https://linkinghub.elsevier.com/retrieve/pii/B9780124080904000025},
	language = {english},
	urldate = {2025-04-25},
	booktitle = {Advances {In} {Atomic}, {Molecular}, and {Optical} {Physics}},
	publisher = {Elsevier},
	author = {Torrontegui, Erik and Ibáñez, Sara and Martínez-Garaot, Sofia and Modugno, Michele and Del Campo, Adolfo and Guéry-Odelin, David and Ruschhaupt, Andreas and Chen, Xi and Muga, Juan Gonzalo},
	year = {2013},
	doi = {10.1016/B978-0-12-408090-4.00002-5},
	pages = {117--169},
}

@article{berry_transitionless_2009,
	title = {Transitionless quantum driving},
	volume = {42},
	issn = {1751-8113, 1751-8121},
	url = {https://iopscience.iop.org/article/10.1088/1751-8113/42/36/365303},
	doi = {10.1088/1751-8113/42/36/365303},
	number = {36},
	urldate = {2025-04-25},
	journal = {Journal of Physics A: Mathematical and Theoretical},
	author = {Berry, M V},
	month = sep,
	year = {2009},
	pages = {365303},
}

@article{torrontegui_fast_2011,
	title = {Fast atomic transport without vibrational heating},
	volume = {83},
	copyright = {http://link.aps.org/licenses/aps-default-license},
	issn = {1050-2947, 1094-1622},
	url = {https://link.aps.org/doi/10.1103/PhysRevA.83.013415},
	doi = {10.1103/PhysRevA.83.013415},
	language = {english},
	number = {1},
	urldate = {2025-04-25},
	journal = {Physical Review A},
	author = {Torrontegui, E. and Ibáñez, S. and Chen, Xi and Ruschhaupt, A. and Guéry-Odelin, D. and Muga, J. G.},
	month = jan,
	year = {2011},
	pages = {013415},
}

@article{plata_taming_2021,
	title = {Taming the {Time} {Evolution} in {Overdamped} {Systems}: {Shortcuts} {Elaborated} from {Fast}-{Forward} and {Time}-{Reversed} {Protocols}},
	volume = {127},
	issn = {0031-9007, 1079-7114},
	shorttitle = {Taming the {Time} {Evolution} in {Overdamped} {Systems}},
	url = {https://link.aps.org/doi/10.1103/PhysRevLett.127.190605},
	doi = {10.1103/PhysRevLett.127.190605},
	language = {english},
	number = {19},
	urldate = {2025-04-25},
	journal = {Physical Review Letters},
	author = {Plata, Carlos A. and Prados, Antonio and Trizac, Emmanuel and Guéry-Odelin, David},
	month = nov,
	year = {2021},
	pages = {190605},
}

@article{raynal_shortcuts_2023,
	title = {Shortcuts to {Equilibrium} with a {Levitated} {Particle} in the {Underdamped} {Regime}},
	volume = {131},
	issn = {0031-9007, 1079-7114},
	url = {https://link.aps.org/doi/10.1103/PhysRevLett.131.087101},
	doi = {10.1103/PhysRevLett.131.087101},
	language = {english},
	number = {8},
	urldate = {2025-04-25},
	journal = {Physical Review Letters},
	author = {Raynal, Damien and De Guillebon, Timothée and Guéry-Odelin, David and Trizac, Emmanuel and Lauret, Jean-Sébastien and Rondin, Loïc},
	month = aug,
	year = {2023},
	pages = {087101},
}

@article{dalfovo_theory_1999,
	title = {Theory of {Bose}-{Einstein} condensation in trapped gases},
	volume = {71},
	copyright = {http://link.aps.org/licenses/aps-default-license},
	issn = {0034-6861, 1539-0756},
	url = {https://link.aps.org/doi/10.1103/RevModPhys.71.463},
	doi = {10.1103/RevModPhys.71.463},
	language = {english},
	number = {3},
	urldate = {2025-04-28},
	journal = {Reviews of Modern Physics},
	author = {Dalfovo, Franco and Giorgini, Stefano and Pitaevskii, Lev P. and Stringari, Sandro},
	month = apr,
	year = {1999},
	pages = {463--512},
}

@article{cirac_goals_2012,
	title = {Goals and opportunities in quantum simulation},
	volume = {8},
	copyright = {http://www.springer.com/tdm},
	issn = {1745-2473, 1745-2481},
	url = {https://www.nature.com/articles/nphys2275},
	doi = {10.1038/nphys2275},
	language = {english},
	number = {4},
	urldate = {2025-04-28},
	journal = {Nature Physics},
	author = {Cirac, J. Ignacio and Zoller, Peter},
	month = apr,
	year = {2012},
	pages = {264--266},
}

@misc{deffner_quantum_2019,
	title = {Quantum {Thermodynamics}: {An} introduction to the thermodynamics of quantum information},
	copyright = {arXiv.org perpetual, non-exclusive license},
	shorttitle = {Quantum {Thermodynamics}},
	url = {https://arxiv.org/abs/1907.01596},
	doi = {10.48550/ARXIV.1907.01596},
	urldate = {2025-04-28},
	publisher = {arXiv},
	author = {Deffner, Sebastian and Campbell, Steve},
	journal = {ArXiV},
	year = {2019},
	note = {Version Number: 1},
}

@article{Lewis1967,
  title = {Classical and Quantum Systems with Time-Dependent Harmonic-Oscillator-Type {Hamiltonians}},
  author = {Lewis, H. R.},
  journal = {Phys. Rev. Lett.},
  volume = {18},
  issue = {13},
  pages = {510--512},
  numpages = {0},
  year = {1967},
  month = {Mar},
  publisher = {American Physical Society},
  doi = {10.1103/PhysRevLett.18.510},
  url = {https://link.aps.org/doi/10.1103/PhysRevLett.18.510}
}

\end{document}